# Behavioural Analytics: Beyond Risk-based MFA

M.T. Dlamini[1,6], J.H.P. Eloff[1,2], HS Venter[1], M.M. Eloff[5], R.P.S. Henha Eyono[3], N.N. Mosola[4]

[1]*ICSA Research Group, Department of Computer Science, University of Pretoria, Pretoria, South Africa*
[1]{mdlamini, eloff, hventer}@cs.up.ac.za

[2]*Cyber-security & Big Data Science Research Group, Department of Computer Science, University of Pretoria, South Africa*
[2]jan.eloff@up.ac.za

[3]*Department of Computer Science, University of Cape Town, Cape Town, South Africa*
[3]royeyono@yahoo.com

[4]*Department of Mathematics and Computer Science, National University of Lesotho, Lesotho*
[4]nmosolan@gmail.com

[5]*Institute for Corporate Citizenship, University of South Africa, Pretoria, South Africa*
[5]eloffmm@unisa.ac.za

[6]*Command, Control and Information Warfare, Defence, Peace, Safety and Security, Council of Scientific and Industrial Research, Pretoria, South Africa*
[6]TDlamini1@csir.co.za

*Abstract*— **This paper investigates how to effectively stop an attacker from using compromised user credentials to gain authorized entry to systems that they are otherwise not authorised to access. The proposed solution extends previous work to move beyond a risk-based multi-factor authentication system. It adds a behavioural analytics component that uses keystroke dynamics to grant or deny users access. Given the increasing number of compromised user credential stores, we make the assumption that criminals already know the user credentials. Hence, to test our solution, users were given authentic user credentials and asked to login to our proof-of-concept. Despite the fact that all illegitimate users in our test cases were given the correct user credentials for legitimate users, none of these were granted access by the system. This demonstrates zero-tolerance to false positives. The results demonstrate the uniqueness of keystroke dynamics and its use to prevent users with stolen credentials from accessing systems they are not authorized to access.**

*Keywords*— **Behavioural analytics, risk-based MFA**

## I. Introduction

For some years now, user credentials have been one of the central mechanisms used to identify and verify that users of systems are exactly who they claim to be. User credentials have always been the main gate keeper for granting or denying users' access to systems. It is not surprising therefore that over the past few years, there has been an increase in the number of attacks targeted at user credentials [15][20][31]. A report by Verizon reflects that user credential breaches have increased from 280 in 2014 to 990 in 2015 [16]. This is an increase of more than 300%. Lawless Research [20] report that in 2015 there were approximately 800 breaches that exposed 169 million user credential records in the United States of America alone. The study by Verizon also argues that breaches which compromised user credentials accounted for 63% of all breaches in 2015 [16]. This is more than half of all breaches discovered in 2015. More recently, some reports were circulating that 272.3 million user credentials have been stolen for Gmail, Hotmail and Yahoo mail [18]. Clearly, user credentials are the target and provide low-hanging fruits to attackers. However, the only mitigation strategy most companies respond with; is to force users to change their passwords and/or a move to enforce strong passwords – passwords with a variable mix of normal and special characters.

Contrary to the move for dynamic and strong passwords, users often prefer to have few rarely changing weak passwords that are easy to remember Alshanketi et al. [17] for usability. Moreover, these are often re-used across multiple systems or services without any alterations [17][19][20]. Phishing and key-logger attacks are responsible for harvesting user credentials on an industrial scale. Hence, reports on a security breach that revealed Facebook's founder to have used 'dadada' as a password for both Twitter and Pinterest accounts are not far-fetched [11] - [14]. Such a weak password could be brute-forced in less than 25 seconds [14]. This might sound like an isolated incident. However, there are more of such cases.

The ever increasing threat of stolen, weak and often re-used credentials has made systems to be vulnerable. Therefore, some researchers on one end argue that user credentials are becoming obsolete [21][28][30][31]. Some argue that despite countless research efforts to replace user credentials; their pervasiveness even today demonstrate their credibility [27] - [29]. The research efforts around user credentials have been impressive. But the results are disappointing [28]. Not only have researchers failed to get rid of user credentials, but they continue to multiply and are still widely used for verifying user authenticity although they are known to insecure. Therefore, some argue that user credentials are like 'salt' ('salt' is good when added in small quantities to something else, but cannot be consumed on its own) [27][32]. Given the pervasiveness of user credentials, it seems plausible to design new authentication systems around their limitations rather than trying to replace them [19][20]. Hence, this paper argues that user credentials on their own are no longer sufficient to secure systems as rightly pointed by other researchers [26][27].

Furthermore, this paper attempts to solve the rising problem of stolen, weak and often re-used user credentials. It answers the question on; how to stop an unauthorized user armed with the compromised credentials of an authorized user from accessing systems they are not authorized to access? This paper builds on a risk-based multi-factor authentication Dlamini et al. [23]; Stinegger et al. [32] and Adam et al. [33] by adding behavioral analytics to authenticate users with other attributes beyond user credentials.

The main contribution of this paper focuses in making use of behavioral analytics like contextual data and keystroke dynamics to authenticate users on top of their credentials. The proposed solution makes authentication decisions to deny or grant access based primarily on user behavior which is derived as the user enters their credentials over a period of time.

The rest of the paper is structured as follows: section 2 discusses related work. Section 3 presents the proposed model. Section 4 discusses the results. Section 5 concludes the paper and provides future work.

## II. RELATED WORK

Despite user credentials' many known weaknesses and vulnerabilities, they continue to dominate authentication mechanisms used by most systems. Potter [28] argues that user credentials are increasingly becoming insecure. However, the alternative solutions do not look promising. Unsurprisingly, the banking sector's automatic teller machines (ATMs) are still reliant on a supposedly random four character personal identification number (PIN) and a plastic card with a chip or magnetic strip which could be cloned to authenticate clients for access to their money [28]. Somehow, the banking industry still believe that four random characters could keep safe people's hard-earned money.

Several researchers have tried to address the problem of stolen or weak user credentials to strengthen authentication systems using different approaches. For example, Kumar et al., [1]; and De Ru and Eloff [26] focus on keystroke dynamics that capture the typing patterns of users. This work classifies users into fast, moderate and slow groups [1]. De Ru and Eloff [26] capture the typing pattern of user credentials and store them in a typing template for use in authentication. However, Kumar et al., [1] and De Ru and Eloff [26] fail to acknowledge that at any particular point and time a user can fall into all three groups. This may depend on other external factors such as;
- the expertise of the current user,
- the knowledge of the target string being used as an authenticator,
- the user's emotional state and,
- the session [6].

Hence, the approach of Kumar et al., [1] and De Ru and Eloff [26] is associated with high false positives and false negative rates (FPR and FNR respectively). Table 1 below shows the difference between the two – i.e. false negatives and false positives. It shows that a false negative occurs when a legitimate user with correct credentials is denied access to a system that they are authorized to access. A false positive occurs when an illegitimate user with the correct user credentials is granted access to a system they are not authorized to access. Furthermore, the clusters therein are created from single key press in a typing session. The work fails to cater for a combination of key presses e.g. shift-A for capital letter A.

The work of Haque et al., [2] authenticates users based on their habitual typing patterns and a target string. The authors use key press events such as dwell time (i.e. key press time), flight time (i.e. inter key press time), and the total time (i.e. session time) to perform classification and verification of users. Similar to Kumar et al., [1], the work of Haque et al., [2] also suffers from high FPR and FNR. For example, an authentic user might be flagged an imposter and denied access for just one incorrect character in the target string. Furthermore, this work fails to consider typographic errors yet this would have a huge impact on the flight and dwell times. For example, the use of delete and backspace keys in case of errors could easily render a legitimate user as an imposter.

TABLE I
FALSE NEGATIVES VS FALSE POSITIVES

|  | Legitimate user with correct credential | Illegitimate user with correct credentials |
|---|---|---|
| **Grant Access** | Allowed | "False Positive" |
| **Deny Access** | "False Negative" | Allowed |

Haque et al., [2] explore the dwell and flight time of a user's typing patterns in a dynamic manner. This work adds a genuine acceptance rate (GAR) to quantify the acceptance of an authentic user. This is based on a user's typing features falling within a certain threshold. Using a maximum likelihood estimation (MLE), the authors claim a 92% success rate, 4% FPRs and 4% FNRs. However, this work Haque et al. [2] does not use the concept of clustering to detect outliers. The outlier detection algorithm computes a value and compares it to just one known list of points. This is not enough and it degrades the integrity of the results.

On the contrary, Ali et al. [6] takes a different dimension and considers the different external factors that affect user's typing behaviour. This work considers time, health conditions, emotional state and environment. However, Ali et al. [6] use statistical methods such as the mean and standard deviations. These often produce good results when given large samples. They fail to provide accurate results when the population size is small. Pisani et al. [10] proposes a system that uses multiple base classifiers (called ensemble) and adaptive algorithms. Hence, Pisani et al. [10] extends the work of Ali et al. [6] and overcome the challenges brought by change of typing behaviour due to time.

Donozo, Ito and Nakakuni [4], propose a multi-modal behaviour analytics model that is based on a Pareto learning self-organizing map (PL SOM). The PL SOM is meant to improve the success rate and reduce FPR and FNR. However, it is difficult to determine a threshold and keep track of each user's login attempt. Moreover, the size of a Pareto optimal set is large when compared to training and learned data. This makes the system to be heavy in memory and processing power usage [4]. Hence, the solution proposed in Dozono et al. [4] cannot be deployed in single-user devices such as mobile devices.

Gurary et al. [5] implement a light-weight solution for mobile devices using keystroke dynamics. This work uses a soft keyboard for a touch screen. It focuses on the duration of a touch and the amount of time between consecutive touches. The authors added motion events instead of key press events. Additionally, Gurary et al. [5] shows some interesting results that are derived from K-Nearest Neighbour (KNN), binary decision trees and naïve Bayes models. These are used for classifying a new login request. The problem with this approach is that group instance points becomes a big cluster. Therefore, a login request from two users could result in a tie, an almost impossible stance for unique behavioural patterns.

Alshanketi et al. [7] looks into the feasibility of using keystroke dynamics in small devices, with limited resources. This study includes both statistical and neural networks techniques. The neural networks' results are better than the results of the statistical methods. Hence, this work shows a significant improvement on the FPRs. The authors use a random forest algorithm to improve the accuracy of the system.

Dholi and Chaudhari [8] also propose the use of a K-NN which shows some improvement on the success rates, FPRs and FNRs. However, this work does not have any outlier detection algorithm to check the neighbouring clusters. Hence, any feature falling out of a potential cluster is immediately flagged as an impostor. Huran et al. [9] deviates from using a K-NN and uses a multi-layer perceptron (MLP) neural network. The MLP seems to work better in rejecting most impostors. However, it requires long training processes for large data sets.

Existing literature makes a plausible attempt at using keystroke dynamics for authentication purposes. However, most of it fails to take into consideration other contextual data like geo-location, access device of login origin. These could be used to further increase success rates, and reduce FPRs and FNRs. The covered literature does not discuss how to mitigate the increasing threat posed by imposters with correct but stolen credentials. Furthermore, it also does not discuss the possibility of denying access and eventually locking out a legitimate user with the correct credentials but unusual keystroke dynamics.

### III. PROPOSED MODEL

Figure 1 below depicts the proposed model. The model is first trained to learn the behavioural patterns of authentic users. A number of features are extracted during the training process and stored in a repository. These include session times, typing speed, keystroke combinations, pressure on the keys etc. The model learns more as users continuously make use of it. All successful attempts are added into the repository for the system to further learn and improve the accuracy of the results. The model has various phases as discussed below.

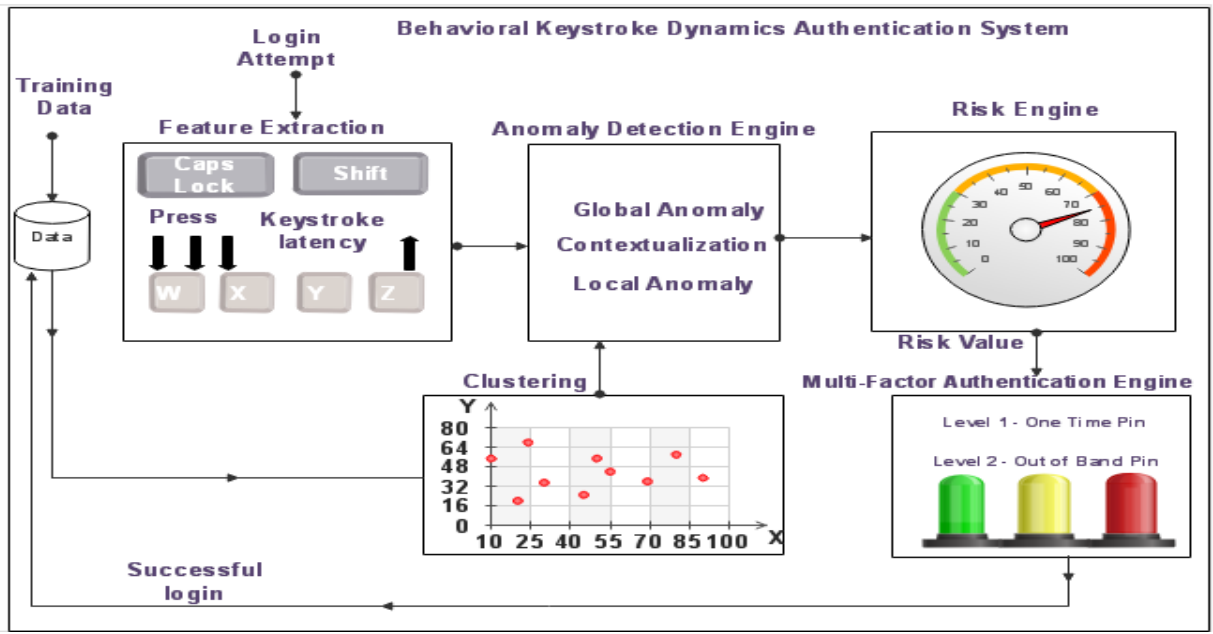

Figure 1: Behavioural Keystroke Dynamics Authentication System model

#### A. Login Attempt

The model uses traditional username and password for authentication. However, it adopts a two-step authentication process. The user types a username. If this is correct, the user subsequently types a password. This two-step authentication is beneficial because it makes it difficult for attackers to crack the system. An attacker will not even get to enter a password if they fail to enter the correct username. This approach could easily become a weakness if one considers an attacker with correct usernames which they know for sure are correct because of the first verification. The attackers' only task would be to take each username and brute-force for the password. The authors acknowledge this limitation. However, the proposed solution requires that both username and password are entered manually without any auto complete. So a brute-force and dictionary attack would be an unsurmountable task to the attacker because they will have to manually type in each password. Furthermore, the more an attacker tries different password combinations; the more the system learns from their typing behaviour that they are actually not the authorized user from the behavioural partner and this way our solution could help profile the attackers before they are kicked out. The model extract keystroke features from the user as they type in their credentials.

#### B. Feature Extraction

As the user types in their credentials on login attempt, the model extracts keystroke features to create a user profile. This happens in real-time in the background. Currently, our model extracts the following features:
- Typing rate
- Timestamp
- Left and Right Shift
- Caps Lock

- Backspace
- Delete
- Geo-location
- Keystroke latency
- Keystroke pressure

Collectively, these features are used to model and profile a user's behavioural patterns. Existing literature seems to favour the use of keystroke latency and keystroke pressure for user authentication purposes. The proposed model also considers the two (i.e. keystroke latency and pressure) of every keystroke pressed is recorded within a login session. It then counts the feature set and measures the number of times a key is pressed. Furthermore, the model also captures and uses other contextual data like geo-location, time-zone etc. If used appropriately, such contextual data can help detect and proactively prevent unauthorized login attempts. All the captured data (i.e. feature extraction and contextual data) is then stored in a data store.

*C. Data Store*

Historical data that consists of successful login attempts is kept in a data store. The model autonomously learns new behavioral patterns and performs self-updates for each and every successful login attempt. The more successful login attempts a user makes, the more accurate the model would correctly classify and detect future login attempts. The historical data is drawn from the database during a login attempt and taken into the clustering process.

*D. Clustering*

Clustering is a good technique especially when dealing with big data. It helps to classify and categorize data according to small clusters. The purpose of the clustering process is to facilitate the work of the anomaly detection engine. The clustering process enables the proposed model to trace a user's behavioral patterns at login. When tracing the behavioral patterns, the model performs anomaly detection on a subset of the data to reduce the computational costs. The model uses a *k-means* clustering algorithm. This process also helps to increase the performance of the anomaly detection engine. The clustering algorithm is as shown below:

1. *Place k points into the space represented by the objects that are being clustered. (These points represent initial group centroids).*
2. *Assign each object to the group that has the closest centroid.*
3. *When all objects have been assigned, recalculate the positions of the k centroids.*
4. *Repeat Steps 2 and 3 until the centroids can no longer move. This produces a separation of the objects into groups from which the metric to be minimized can be calculated.*

There are various heuristics used to determine the value of *k* in the *k*-means algorithm. These include choosing *k* by visualisation or choosing *k* within a range or set. We picked the elbow method as our heuristic of choice, due to it being one of the tried and tested method for determining *k* [27].

*E. Anomaly Detection Engine*

The anomaly detection engine takes input from a login attempt and clustered user data. It then assigns a risk level to the user. The anomaly detection part of the model comprises of three stages. These are global anomaly check, contextualization and local anomaly check. The next sub-section discusses each of these stages in detail.

*1) Global anomaly check:* In the global anomaly check, the model checks for login attempts that are considered extreme anomalies. For example, an extreme anomaly would occur when a user who habitually uses a caps lock for capital letters but then starts to use a combination of shift and a letter. In the case that a user fails this global anomaly check, it would mean that their current login attempt deviates from their behavioural patterns. The user is then flagged as high risk. This is achieved by re-clustering historical data points from the current login attempt. The motivation behind re-clustering is that most clustering algorithms, specifically the *k-means*, are sensitive to outliers [24].

In the case that there exists an extreme anomaly point in the dataset, it is often placed in its own cluster. A login attempt being placed in its own cluster is a clear indication of an anomaly. All users that fail the global anomaly check are flagged as high risk. This does not necessarily imply that such users would be denied access. However, it means that extra caution must be given before they are granted access. This is enforced by the MFA engine which authenticates users based on their risk profile. Should it so happen that the global anomaly check is passed, the model moves to contextualization.

*2) Contextualization:* This phase contextualizes the observed login attempt for further analysis. It is implemented using a nearest centroid algorithm. This algorithm assesses and classifies each of the centroids of the *k* clusters to determine one that is most similar to the observed login attempt. Similarity is measured using an Euclidean distance as follows. Firstly, the model selects a cluster which best represents the observed login attempt. It then deduces whether the observed login attempt is an anomaly or not. The model then moves to determine the context of the login attempt. Once this is found, it is then fed into the stage on anomaly detection engine i.e. local anomaly check.

3) *Local anomaly check:* The local anomaly check performs further analysis using classical statistical techniques. This assigns a risk-level to the observed login attempt. The local anomaly check assesses whether a login attempt is an outlier or not. The first stage of the local anomaly check is focused on representing the respective distances between each member of the contextual cluster and its centroid.

This entails retrieving the distances, arranging them in ascending order. When retrieving the distances and arranging them, it is important to avoid redundancy. This is because emphasis is placed on the distance value that is represented and not necessarily on how many times it appears.

In the case that the Euclidean distance between the observed login attempt and the centroid is strictly greater than two specific thresholds, then the login attempt is flagged as an outlier. More specifically, if it is strictly greater than the first but less than the second threshold, the model refers to it as a first-degree outlier. However, if it is greater than or equal to the second threshold, it is referred to as a second-degree outlier.

For example, failing a global anomaly check is considered as a second-degree outlier and it passes control to the MFA engine.

*F. Multi-Factor Authentication Engine*

The MFA engine assesses the authenticity and risk level of a user's login attempt and applies the corresponding level of authentication measure. The MFA engine acts as a barrier of defence. Furthermore, it also allows for the model to learn from behavioral patterns. The engine has different levels of authentication measures, each corresponding to the risk-level of a user. The model uses following MFA mechanisms for the different risk-levels:
1. First-Degree Outlier: One Time Password (OTP)
2. Second-Degree Outlier: Out-of-Band Authentication (OOB)

A first-degree outlier prompts the MFA engine to issue an OTP as an extra layer of defence on top of the correct user credentials. Once prompted for an OTP, the model then sends a temporary password to a remote device, more often, a cell-phone or an email address.

A second-degree outlier prompts the MFA engine to issue an OOB. An OOB is a relatively new authentication technique compared to OTP. OOB redirects authentication to a remote device. A user would for example be re-directed to authenticate on a different device. Once a user is authenticated; control is then passed back to the model. OOBs can help protect against man-in-the-browser attacks and many others.

## IV. RESULTS

The experiment started by training the proposed model with login credentials of a group of authentic and authorized users. The users were asked to enter the login credentials for a number of rounds in order to train the model. The system made use of the session duration time and the typing speed to illustrate and demonstrate the practicability of the model. This is meant as a proof-of-concept experiment. This process started with a training set of instances clustered over a number of clusters. This is the data that goes into the data store. The model was then tested on a small sample of users to illustrate the point and its practicability. Our sample size of ten testing users is not in any way significant to draw conclusive results.

Each of the users was given the login details of an authentic user on which the model was trained on initially. However, all login attempts of these unauthorized users' with correct credential were flagged as anomalies and further denied access on their first attempt. Our solution only granted access to authentic users. Of the all the testing users, sixty percent failed the global anomaly check. The use of a mix of characters and numbers in the password field made it difficult for the users to traverse through the keyboard. This worked to our advantage. All the users were made aware of the extra barrier of security that uses behavioural analytics of keystrokes. However, they still were denied access. Figure 2 above illustrates an example of a login attempt that fails the global anomaly check.

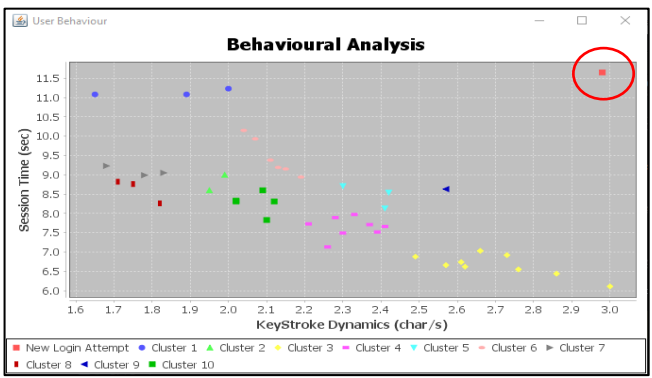

Figure 2: Failed Global Anomaly check

This particular user's login attempt is shown at the right top corner of the x-y plot inside the red circle. This is too far from any of the clusters. Such login attempts require stringent MFA for they carry the highest risk. They trigger an OOB from the MFA.

40 percent of the users that failed the local anomaly check. Even though these users' login attempts failed to comply with correct behavioural patterns, they were not necessarily flagged as extreme anomalies. Figure 3 below illustrates a login attempt that failed the local anomaly check.

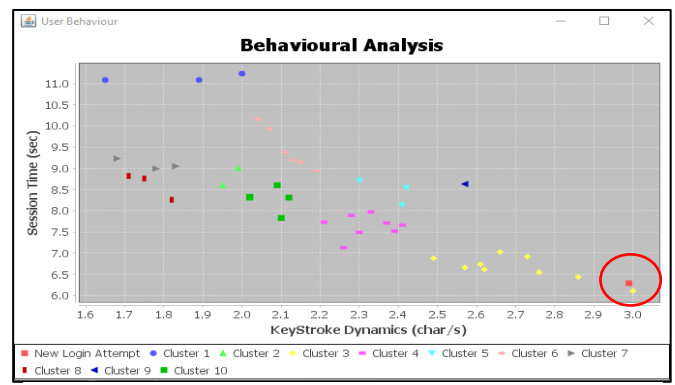

Figure 3: Failed Local Anomaly check

In this figure, the cluster of context is cluster 1 and the centroid covers approximately a radius of 0.14. However, the user's login attempt maintained a distance of 0.26 in the red circle. Existing models that uses behavioural keystroke dynamics would classify such an anomaly as a normal pattern. This is not the case with our model, as this is classified as an outlier.

From the results, we can conclude that the proposed solution provides a plausible solution to solve the challenges posed by attackers using compromised user credentials to access systems they are not authorised to access. The model achieved its goal and successfully prevented unauthorised users with the correct user credentials from gaining access. The next section concludes the paper and provides directions for future work.

## V. CONCLUSION AND FUTURE WORK

Most user credentials are already compromised or they soon will be. This paper illustrated the use of behavioral keystroke analytics to effectively deny unauthorized users with correct user credentials from accessing systems that they are not authorized to use. The proposed model proved to be effective and resilient. It denied access for all the unauthorized users without necessarily flagging for false positives or negatives. The results demonstrate the uniqueness of keystroke dynamics. In future, we plan to use continuous authentication to counter session hijacks. For performance issues, we will consider other measures of calculating the Euclidean distance. For robustness and improved accuracy; the solution needs more experiments with a large sample size of the training data of at least ten legitimate users instead of one and even more testing users.

**Moses Dlamini** is PHD student at the University of Pretoria. He is also a researcher in the DPSS unit at the CSIR.

**Jan Eloff** is the Deputy Dean Research & Postgraduate: Faculty of Eng., Built Environment and a full professor in Computer Science at the University of Pretoria. He leads the Big Data Science and Cyber-security research group.

**Hein Venter** is a professor at the University of Pretoria. He is a research group leader for Information and Computer Security Architectures (ICSA).

**Mariki Eloff** is a full professor at the University of South Africa. She is also a chief researcher at UNISA's Institute of Corporate Citizenship.

**Roy Henha Eyono** is a BSc Honors student at the University of Cape Town. He works at the CSIR during his vacations.

**Napo Mosola** is a lecturer at the National University of Lesotho. He has just completed his MSc in Computer Science focusing of crypto.